\documentclass[amsmath,amssymb,preprint,showpacs]{revtex4}

\usepackage{graphicx}
\usepackage[usenames]{color}

\begin{document}

\title{Electronic correlations in iron-pnictide superconductors and beyond; what can we learn from optics}
\author{L. Degiorgi$^1$} \affiliation{$^1$Laboratorium f\"ur
Festk\"orperphysik, ETH - Z\"urich, CH-8093 Z\"urich,
Switzerland}

\date{\today}

\begin{abstract}
The Coulomb repulsion, impeding electrons' motion, has an important impact on the charge dynamics. It mainly causes a reduction of the effective metallic Drude weight (proportional to the so-called optical kinetic energy), encountered in the optical conductivity, with respect to the expectation within the nearly-free electron limit (defining the so-called band kinetic energy), as evinced from band-structure theory. In principle, the ratio between the optical and band kinetic energy allows defining the degree of electronic correlations. Through spectral weight arguments based on the excitation spectrum, we provide an experimental tool, free from any theoretical or band-structure based assumptions, in order to estimate the degree of electronic correlations in several systems. We first address the novel iron-pnictide superconductors, which serve to set the stage for our approach. We then revisit a large variety of materials, ranging from superconductors, to Kondo-like systems as well as materials close to the Mott-insulating state. As comparison we also tackle materials, where the electron-phonon coupling dominates. We establish a direct relationship between the strength of interaction and the resulting reduction of the optical kinetic energy of the itinerant charge carriers.

\end{abstract}


%\keywords{rare-earth,tellurides}

\maketitle

\section{Introduction}
Since the discovery of superconductivity in the iron-pnictide materials \cite{kamihara,rotter,chen,sefat}, the issue about the strength of their electronic correlations (i.e., the repulsive interactions among electrons) has been widely debated and it is still matter of intense investigations. This issue is of stringent importance for several families of (high temperature) superconductors, since superconductivity transition temperatures beyond 20 K cannot be yielded within conventional mechanism based on electron-phonon coupling \cite{bcs}. In this respect, an ample discussion was also generated on whether the novel family of iron-pnictides does share common features and similarities with the high temperature superconducting cuprates (HTC), discovered more than 20 years ago \cite{mueller}. In the cuprates electronic correlations are so strong that the parent compounds, out of which superconductivity originates, are Mott insulator. On the other hand, there is mounting evidence that the parent compounds of the iron-pnictide superconductors are bad metals, where electronic correlations are sufficiently strong to place them close to the boundary between itinerancy and interaction-induced electronic localizations \cite{si}. Moreover in both families, superconductivity develops when magnetism, characterizing in part their phase diagram, is destroyed by dopings. This led to the conjecture that exchange of magnetic fluctuations may provide the glue, binding electrons into Cooper pairs \cite{mazin}.

The degree of electronic correlations has a direct impact on the charge dynamics, generally evinced from the optical conductivity \cite{si,basovreview}. Thanks to the development of appropriate optical methods, this latter quantity can be achieved nowadays with great precision over an extremely broad spectral range, a prerequisite of paramount importance in order to track the implications of electronic correlations. Since electronic correlations significantly impede the mobility of the electrons, they consequently lead to a substantial reduction of the kinetic energy of the itinerant charge carriers with respect to the expectation for nearly-free or non-interacting particles. Precisely, one has to distinguish among two cases; namely, quenching the Drude weight by local (Hubbard-like) correlations or by interactions with a bosonic mode (i.e., phonons or spin fluctuations) \cite{benfattocomment}. The sum rule for the first case would reveal a spectral weight change from $n/m_{LDA}$ ($m_{LDA}$ being the crude mass for the local-density approximation (LDA)) to $n/m_{U}$ (with $m_U$ generally bigger than $m_{LDA}$). For the second case, the sum rule would still lead to a total weight given by $n/m_{LDA}$, the latter being then redistributed between a Drude peak with weight proportional to $n/m^*$ ($m^*$ is the renormalized mass by the bosonic mode and is greater than $m_{LDA}$) and a so-called incoherent part encountering the remaining weight.

Qazilbash et al. recently made an interesting survey of the electronic correlations by looking to a wealth of materials, ranging from conventional metals to Mott insulators \cite{basov}. Along the same line of arguments previously introduced by Millis et al. (Ref. \onlinecite{millis}), they proposed a quantitative approach for the calculation of the ratio between the optical kinetic energy ($K_{opt}$) and the band kinetic energy ($K_{band}$). The former is obtained from the integral of the effective (Drude) metallic component of the optical conductivity (also referred as the coherent part of the single electron excitations in an interacting metallic system), while the latter quantity (also known as the kinetic energy of the underlying non-interacting system) is extracted form $ab$ $initio$ (tight-binding) band-structure calculations neglecting the electron-electron interactions \cite{si}. Their analysis establishes a regime of moderate correlations for the iron-pnictide superconductors and leads to a nice correspondence between the degree of electronic correlations and the reduction of the empirical kinetic energy of the charge carriers compared with theory for a large variety of materials. Materials, for which the electron-phonon coupling predominately shapes the intrinsic physical properties, do not show, on the other hand, a significant reduction of the electrons' kinetic energy (Fig. 3 in Ref. \onlinecite{basov}).

Motivated by the work reported in Ref. \onlinecite{basov}, we undertook a systematic optical investigation of the Co-doped BaFe$_2$As$_2$ family, spanning the entire phase diagram \cite{lucarelli}. We proposed to establish the ratio $K_{opt}/K_{band}$, exclusively from spectral weight arguments based on the experimental findings. %We assumed that $K_{opt}/K_{band}$ is in principle equivalent to the ratio between the effective Drude weight and the total spectral weight encountered in the optical conductivity from zero up to a cut-off frequency at the onset of the electronic interband transition. We give clear-cut evidence for moderate electronic correlations between the parent compound ($x$) and the optimal doping ($x$=0.061), and for a crossover to a regime of weak interacting and nearly-free electron metals for the overdoped ($x\le$0.11) Ba(Co$_x$Fe$_{1-x}$)$_2$As$_2$ materials \cite{lucarelli}. 
The astonishing good correspondence of the degree of electronic correlations with the evolution of the superconducting phase in Ba(Co$_x$Fe$_{1-x}$)$_2$As$_2$ (i.e., the superconducting dome defined by the $T_c$ values in the phase diagram \cite{chu}), makes us confident that one can expand our procedure for other materials. The goal of this communication is to review our estimation of the electronic correlations across the phase diagram of the Ba(Co$_x$Fe$_{1-x}$)$_2$As$_2$ compounds and place it in a broader context by revisiting from a similar perspective other variably correlated materials. We will compare the iron-pnictides with other superconductors, like the HTCs and the boron-carbides and fullerenes, and with strongly correlated $f$- and $d$-electron systems (Kondo and heavy electron materials), as well as linear chain organic Bechgaard salts. Our review will be complemented with representative materials characterized by a CDW ground state, for which the electron-phonon coupling should dominate.

We show that an appropriate spectral-weight analysis can indeed reliably reveal the degree of electronic correlations and thus provide a valuable and exclusive experimental instruments, free from any $ad$ $hoc$ theoretical constrains, in order to discriminate among various systems with different levels of correlations. We principally analyze the real part ($\sigma_1(\omega)$) of the optical conductivity which is mainly obtained by first measuring the reflectivity ($R(\omega)$) over an extremely broad spectral range. This permits to perform reliable Kramers-Kronig transformation from which we calculate the phase of the complex reflectance and then all optical functions \cite{wooten,grunerbook}.

\begin{figure}[!tb]
\center
\includegraphics[width=12cm]{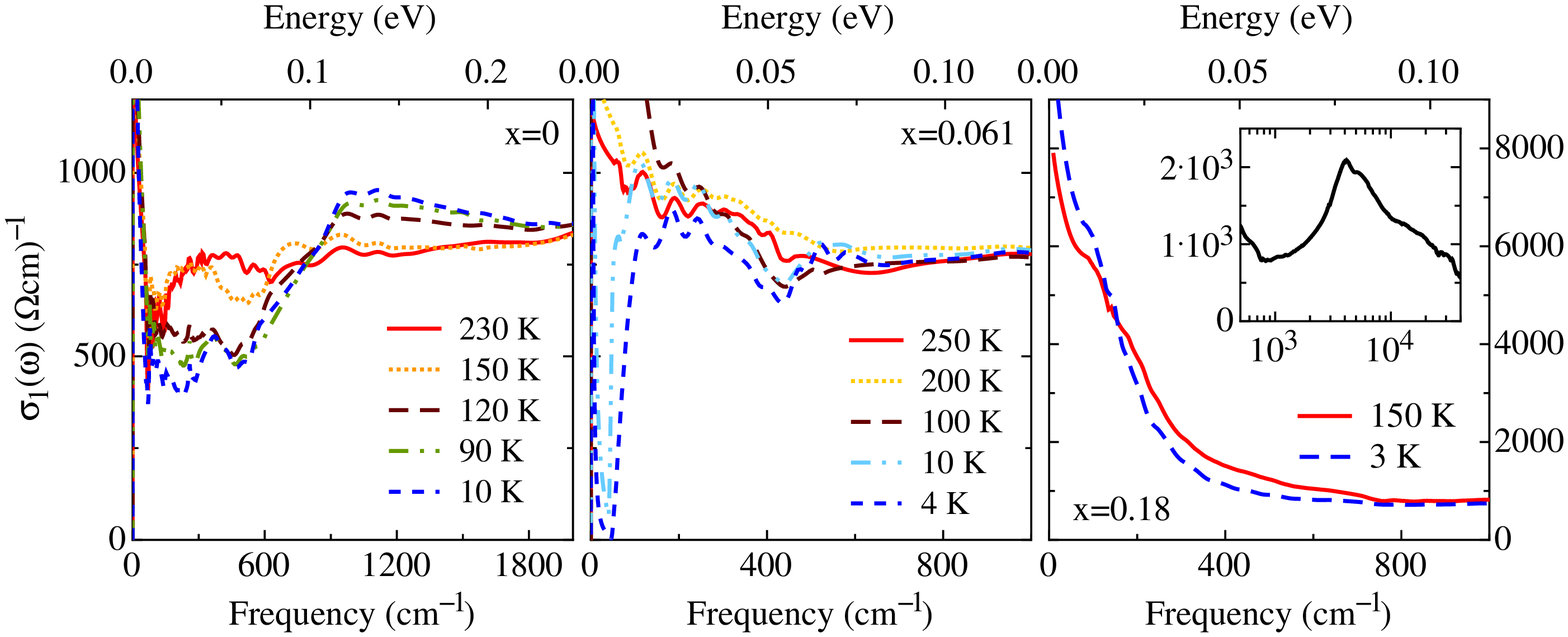}
\caption{(color online) Real part $\sigma_1(\omega)$ of the optical conductivity for Ba(Co$_x$Fe$_{1-x}$)$_2$As$_2$ for $x=$ 0, 6.1\% and 18\% in the far and mid-infrared spectral range at selected temperatures above and below the various phase transitions. The inset displays $\sigma_1(\omega)$ at 300 K for $x$=18\%, emphasizing its representative shape for all compounds at high frequencies up to the ultraviolet \cite{lucarelli}.} \label{sigmapnictides}
\end{figure}

\section{Experimental results and Discussion}
We start our survey by summarizing the most pertinent results on Co-doped BaFe$_2$As$_2$, reported in Ref. \onlinecite{lucarelli} and which serve as an interesting case in point. Figure 1 shows $\sigma_1(\omega)$ in the far- and mid-infrared energy interval for selected Co-dopings (non-superconducting parent compound, optimal doping and non-superconducting overdoped compound) as a function of temperature. The parent compound ($x$=0) clearly displays the opening of the spin-density-wave (SDW)-pseudogap at $T_{SDW}$, while the optimally doped one ($x$=6.1\%) gives clear-cut evidence for the development of the superconducting gap below $T_c$. Finally, the compound ($x$=18\%) at the opposite end of the superconducting dome is in the metallic phase at all temperatures. Overall, there are three energy intervals characterizing $\sigma_1(\omega)$ for all Co-dopings: the effective metallic contribution at low frequencies, a rather flat mid-infrared (MIR) region covering the energy interval between 500 and 1500 cm$^{-1}$ (from now on called MIR band) and the electronic interband transitions with onset at about 2000 cm$^{-1}$ and peaked at 5000 cm$^{-1}$. The metallic part as well as the MIR band turn out to experience the strongest temperature dependence at $T_c$ and/or $T_{SDW}$, while the high frequency excitations (inset Fig. 1) are temperature independent \cite{lucarelli}. In order to account for the various contributions to the excitation spectrum, we applied the well-established phenomenological Drude-Lorentz approach \cite{wooten,grunerbook}. Besides high frequency Lorentz h.o.'s for the interband transitions with onset at $\omega\le$2000 cm$^{-1}$ (inset Fig. 1), $\sigma_1(\omega)$ can be reproduced in great details by adding two Drude terms for the effective metallic contribution and a broad h.o. for the MIR energy interval (inset of Fig. 2) \cite{lucarelli}. Consistent with Ref. \onlinecite{wu}, it turns out that one Drude term is rather narrow, while the second broad one acts as a background to the optical conductivity. These latter phenomenological components fully describe the temperature dependence of $\sigma_1(\omega)$ for all dopings ($x$). The two (narrow and broad) Drude terms imply the existence of two electronic subsystems. In passing, we remark that the approach is too phenomenological to allow speculations on whether the two Drude components have specific orbital character.

\begin{figure}
\center
\includegraphics[width=12cm]{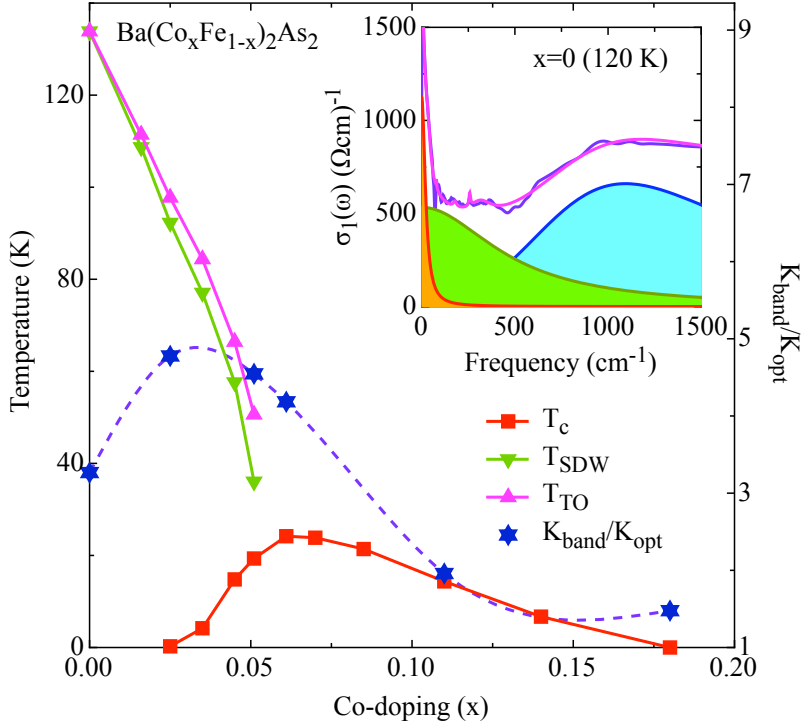}
\caption{(color online) Phase diagram of Ba(Co$_x$Fe$_{1-x}$)$_2$As$_2$, reproduced from Ref. \onlinecite{chu} (left y-axis), and the average of the ratio $K_{band}/K_{opt}$ (eq. (1)) calculated at high and low temperature (right y-axis). $T_c$, $T_{SDW}$ and $T_{TO}$ are the critical temperatures for the superconducting and SDW phase transition as well as for the tetragonal-orthorhombic structural transition, respectively \cite{chu}. All data-interpolation are spline lines as guide to the eyes. The inset displays the optical conductivity at 120 K for the parent compound ($x$=0) with the total Drude-Lorentz fit and its low frequency components; i.e., the narrow and broad Drude terms as well as the mid-infrared (MIR) h.o. (see text). The shaded areas emphasize the respective spectral weights ($\int\sigma_{1N}(\omega)d\omega, \int\sigma_{1B}(\omega)d\omega$ and $\int\sigma_1^{MIR}(\omega)d\omega$, eq. (1)) \cite{lucarelli}.} \label{phase diagram}
\end{figure}

We proposed a scenario where the conduction band derives from $d$-states and splits into two parts: a purely itinerant one close to the Fermi level and represented by the two Drude components as well as by a bottom part with states below the mobility edge and thus rather localized \cite{lucarelli}. This latter part gives rise to the MIR band in $\sigma_1(\omega)$, which turns out to be strongly temperature dependent upon magnetic ordering and affected by the opening of the SDW gap. In order to shed light on the relative distribution of spectral weight among the various (metallic and MIR) components of $\sigma_1(\omega)$, we have proposed to further exploit our phenomenological Drude-Lorentz description by calculating the spectral weight ratio:
{\setlength\arraycolsep{2pt}
\begin{eqnarray}
K_{opt}/K_{band}=\frac{\int\sigma_{1N}(\omega)d\omega+\int\sigma_{1B}(\omega)d\omega}{\int\sigma_{1N}(\omega)d\omega+\int\sigma_{1B}(\omega)d\omega+\int\sigma_1^{MIR}(\omega)d\omega},
\end{eqnarray}}where $\sigma_{1N}(\omega)$, $\sigma_{1B}(\omega)$ and $\sigma_1^{MIR}(\omega)$ are the fit components of the optical conductivity due to the narrow and broad Drude terms, and to the MIR band (inset of Fig. 2), respectively \cite{lucarelli}. Equation (1) thus represents the ratio between the spectral weight encountered in $\sigma_1(\omega)$ in the Drude components ($K_{opt}$) and the total spectral weight collected in $\sigma_1(\omega)$ up to the onset of the electronic interband transitions (i.e., Drude components together with the MIR absorption feature, $K_{band}$). By assuming the conservation of the total charge carriers density, eq. (1) is in principle proportional to the ratio between the effective mass at low energy scales, which includes all renormalizations of local character or due to spin fluctuations, and at high energies, which only includes local effects of Hubbard type \cite{benfatto1}. Therefore, the reduction of $K_{opt}$ with respect to $K_{band}$ (eq.(1)) would solely derive from the effective mass renormalization. Obviously in a multiband scenario as it applies in the iron-pnictides, it could well be that the rearrangement of the total number of charge carriers between hole and electron bands is such that the total charge carriers density is not conserved. In this case, the spectral weight reduction in the effective metallic contribution of $\sigma_1(\omega)$ might not be due to the effective mass (i.e., correlation effects) alone \cite{benfatto2}. One should then carefully address the role of electron-electron interactions on both effective mass and charge carriers concentration. This is beyond the scope of the present discussion. As already emphasized above, our definition of $K_{band}$ may be equivalent to the LDA expectation in the case that local Hubbard-like effects are negligible. Nevertheless, by assuming that the Hubbard renormalization factor would equally affect both $K_{opt}$ and $K_{band}$, we are confident that eq. (1) is an alternative estimation, exclusively obtained from the experimental findings, of the ratio between the optical kinetic energy extracted by integrating $\sigma_1(\omega)$ up to the onset of the electronic interband transitions \cite{basov} and the band kinetic energy extracted from the band structure within the tight-binding approach. We are also aware of the fact that the MIR band may be also contaminated by coherent contributions originating from transitions between bands close to the Fermi level (in the case of the iron-pnictides of hole-hole type) or by the low energy tail contribution of high energy interband transitions. We assume those effects to be small. We have calculated $K_{opt}/K_{band}$ after eq. (1) at high and low temperature, yet still corresponding to the normal or SDW state \cite{commentcorr}. Our $K_{opt}/K_{band}$ for $x$=0 is consistent with the value reported in Fig. 3 of Ref. \onlinecite{basov} for the same compound, therefore reinforcing the validity of our spectral weight arguments. The inverse of $K_{opt}/K_{band}$, which then defines the degree of electronic correlations, is plotted in Fig. 2 as average of the $K_{band}/K_{opt}$ values at low and high temperature \cite{commentcorr} within the phase diagram of the Co-doped 122 iron-pnictides \cite{chu}. $K_{band}/K_{opt}$ thus tracks the evolution of its superconducting dome. Interestingly enough, electronic correlations seem to be stronger for the parent-compound and for Co-dopings $x\le$0.061 than for those in the overdoped range. There is indeed evidence for a crossover from a regime of moderate correlations for $x\le$0.061 to a nearly-free and non-interacting electron gas system for $x\ge$0.11 \cite{lucarelli}.

\begin{figure}[!tb]
\center
\includegraphics[width=12cm]{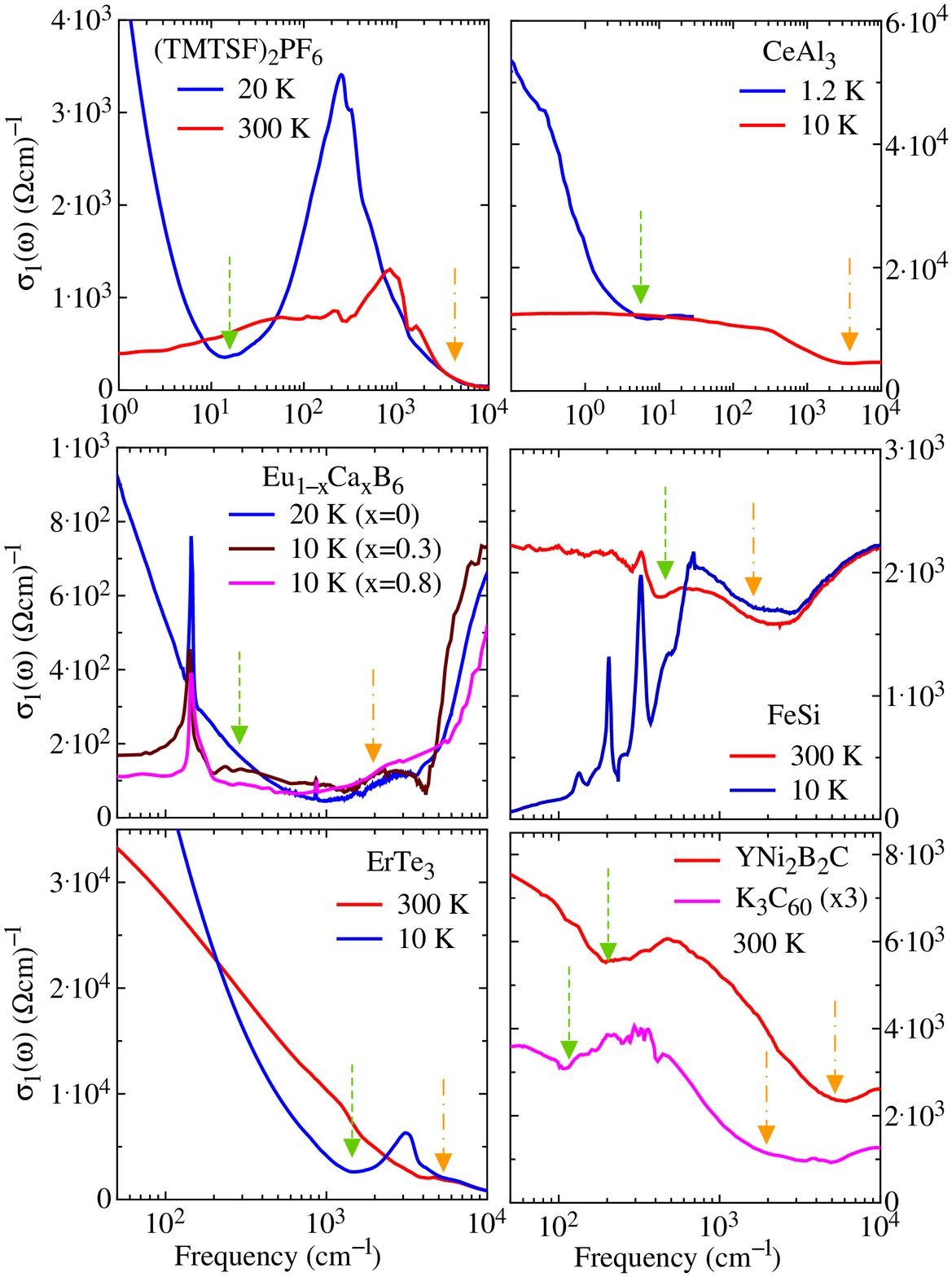}
\caption{(color online) Real part $\sigma_1(\omega)$ of the optical conductivity at selected temperatures for the linear chain Bechgaard salt (TMTSF)$_2$PF$_6$ \cite{schwartz,vescoli}, the heavy-electron CeAl$_3$ \cite{awasthi}, the ferromagnetic metal Eu$_{1-x}$Ca$_x$B$_6$ for $x$=0, 0.3 and 0.8 \cite{degiorgiprl,caimiprl}, the Kondo insulator FeSi \cite{degiorgiepl}, the CDW rare-earth tritellurides ErTe$_3$ \cite{pfuner} and the superconducting boroncarbide YNi$_2$B$_2$C \cite{degiorgiboron} and K$_3$C$_{60}$ \cite{degiorgic60} (for clarity $\sigma_1(\omega)$ has been here multiplied by a factor of three). Please note the different frequency scales for (TMTSF)$_2$PF$_6$ and CeAl$_3$, which extend into the microwave spectral range. The dashed and dashed-dotted arrows mark the cut-off frequencies $\omega_{opt}$ and $\omega_{band}$ (see text) of the integrated optical conductivity leading to $K_{opt}$ and $K_{band}$, respectively \cite{si,basov,commentwband,commentK}.} \label{sigma}
\end{figure}

With the goal to further generalize our approach towards the estimation of $K_{opt}/K_{band}$ based on the spectral weight analysis of $\sigma_1(\omega)$ (eq. (1)), we recollect first of all the electrodynamic response of two other families of superconductors: the alkali-doped fullerenes ($A$C$_{60}$, $A$= K, Rb and Cs) and the boroncarbides $L$Ni$_2$B$_2$C ($L$=Lu, Tm, Er, Ho, Dy and Y). Their electrodynamic response \cite{degiorgic60,degiorgiboron} in the normal state shares general features with $\sigma_1(\omega)$ of the HTCs (see below) \cite{basovreview} and iron-pnictides, as well. The normal state $\sigma_1(\omega)$ of YNi$_2$B$_2$C and K$_3$C$_{60}$ is reproduced in Fig. 3. There is evidence for a two component picture: a Drude response up to $\omega_{opt}$, getting narrow with decreasing temperature, and a mid-infrared band, extending up to $\omega_{band}$ \cite{commentwband} and being reminiscent  of a pseudogap-like excitation. Only about 10\% of the total weight (i.e., $K_{opt}/K_{band}\sim$0.1, \cite{commentK}), encountered in $\sigma_1(\omega)$ up to the onset of the electronic interband transitions, can be ascribed to the itinerant charge carriers. This would imply that both the superconducting fullerenes and boroncarbides are strongly correlated materials at the verge of a Mott insulating state. $K_{opt}/K_{band}$ for both systems are summarized in our figure of merit (Fig. 4), allowing a first comparison with values previously achieved for the iron-pnictides (Fig. 2).

It is now compelling to compare the degree of electronic correlations between these latter families of superconductors and the high-temperature copper-oxides. For the purpose of our discussion, we consider the partial, yet representative series of the underdoped Bi$_2$Sr$_2$CaCu$_2$O$_{8+\delta}$ (Bi2212 (UD), $T_c$= 67 K), the optimally doped YBa$_2$Cu$_3$O$_x$ (Y123 (OP), $T_c$=93.5 K) and the overdoped Tl$_2$Sr$_2$CuO$_{6+\delta}$ (Tl2201 (OD), $T_c$=23 K) compounds \cite{puchkov,basovreview}. The integral of $\sigma_1(\omega)$ up to the onset of the electronic interband transitions (at about $\omega_{band}\sim$1 eV) leads to the total spectral weight proportional to $K_{band}$ after eq. (1). The effective Drude weight, which then defines our $K_{opt}$, is obtained by integrating $\sigma_1(\omega)$ up to a cut-off frequency $\omega_{opt}\sim$ 1000 cm$^{-1}$, thus at the onset of the pseudogap excitation \cite{commentK,puchkov,commentwband}. The ratio $K_{opt}/K_{band}$ is summarized in Fig. 4 for the three chosen compounds and results to be fully in trend with our arguments, proposed above for the iron-pnictides and other superconductors. Our results are even in fair agreement with a semi-theoretical approach on the electron and hole doped cuprates, reported in Refs. \onlinecite{basovreview} and \onlinecite{millis}. It is worth noting, that $\sigma_1(\omega)$ in the non-superconducting La$_{2-x}$Sr$_x$CuO$_4$ ($x$=0.26) \cite{lucarelliHTC} is characterized by a simple Drude response without any signature of the pseudogap excitation, thus implying $K_{opt}/K_{band}\sim$1 (eq. (1)). This is fully consistent with a nearly-free electron limit, as expected for conventional metals like Au or Al \cite{grunerbook,commentmetal}.

The linear chain organic Bechgaard salts \cite{jerome} were also intensively investigated, because of their rich phase diagram. At low temperatures, they provide an interesting arena in which to study the pressure (chemical and applied) induced crossover between a spin-Peierls, a SDW and a superconducting state. At high temperatures one can study the dimensionality-driven evolution between a Mott insulating and an incipient Fermi liquid state upon compressing the lattice. We have shown that the optical conductivity in the metallic state of (TMTSF)$_2$X (where X= SF$_6$, AsF$_6$ or ClO$_4$) is highly anisotropic and display dramatic deviations from a simple Drude response \cite{schwartz,vescoli}. Figure 3 displays, as an example, $\sigma_1(\omega)$ for X=PF$_6$ at 300 and 20 K. One can appreciate the emergence of two prominent features with decreasing temperature: a narrow mode at zero frequency, with a small amount of spectral weight, and a mode centered around 200 cm$^{-1}$, with nearly all the spectral weight expected for the relevant number of carriers and single particle band mass (Fig. 3). We argued that these features are characteristic of a quasi one-dimensional half- or quarter-filled band with Coulomb correlations, and ascribed the finite-energy mode to the correlation (pesudo)gap. We proposed a scenario consistent with a doped Mott-semiconducting behavior of the organic Bechgaard salts \cite{schwartz,vescoli}. The interchain coupling, represented by the transverse charge transfer integral ($t_{\perp}$) between the chains, induces indeed a self-doping of the Mott insulating state of the strict one-dimensional limit (i.e., $t_{\perp}$=0), thus populating the upper Hubbard band and leading to the tiny Drude weight encountered in the zero energy resonance of $\sigma_1(\omega)$ (Fig. 3). $K_{opt}/K_{band}$ corresponds here to the ratio between the spectral weight of the zero energy mode for $\omega<\omega_{opt}\sim$ 20 cm$^{-1}$ (i.e., about 1\% of the total weight) and the total weight of both the zero- and finite-energy modes, up to $\omega_{band}\sim$ 4000 cm$^{-1}$ \cite{commentwband}. This results in $K_{opt}/K_{band}\sim$0.008 (Fig. 4) \cite{commentK}, which is again consistent with the Mott-limit.

Another interesting class of correlated systems is the family of heavy-electron (HE) and more generally Kondo-like materials. Electron-electron interaction leads to two characteristic excitations: a renormalized Drude response and a mid-infrared peak \cite{degiorgirmp,degiorgimagnetic}. The latter originates from a dynamical, correlation-induced gap, as evinced from a many body theoretical approach based on the periodic Anderson model \cite{degiorgiepj}. At low temperatures, it can be viewed as optical gap between two renormalized quasi-particle bands. Paramount examples of HEs are CeAl$_3$, CePd$_3$ and UPt$_3$, just to name a few. Figure 3 displays $\sigma_1(\omega)$ above and below the Kondo temperature $T_K$ for CeAl$_3$ \cite{awasthi}. The spectrum above $T_K$ is compatible with a simple Drude response up to $\omega_{band}\sim$ 4000 cm$^{-1}$ and can be used to estimate $K_{band}$ in eq. (1), while $\sigma_1(\omega)$ below $T_K$ is characterized by the narrow Drude contribution up to $\omega_{opt}\sim$ 60 cm$^{-1}$, due to the effective mass renormalization \cite{commentwband}. The spectral weight of such a renormalized Drude response defines $K_{opt}$. $K_{opt}/K_{band}$ in prototype HEs oscillates between 0.003 and 0.02 \cite{commentK}; i.e., only about 0.3 to 2 \% of the expected total spectral weight in the nearly-free electron or simple band-metal limit is effectively encountered in the (narrow) Drude response of $\sigma_1(\omega)$ \cite{degiorgirmp,degiorgimagnetic,degiorgiepj,awasthi}. These values of $K_{opt}/K_{band}$ also place the HE systems within the strongly correlated regime, comparable to the situation of the Mott-like insulating state (Fig. 4). It is worth noting that our experimental estimation of $K_{opt}/K_{band}$ is particularly suitable in this context. In fact, it is not plagued by the annoying role of the localized $f$-electrons, which would make $K_{band}$ from band-structure calculations less trustable and poorly reliable.

Among the highly correlated electron systems, various rare-earth compounds known as hybridization-gap semiconductors or Kondo insulators \cite{aeppli,degiorgirmp} have also attracted considerable interest recently. The cubic compound FeSi is a prominent example, particularly suitable for investigating aspects of the electronic properties of a $d$-transition metal system that might be related with features of correlation effects in $f$-electron materials. At low frequencies, an Anderson-Mott localization behavior was established from the optical conductivity, while at high frequencies the excitation spectrum resembles that of a conventional semiconductor \cite{degiorgiepl}. Figure 3 displays $\sigma_1(\omega)$ of FeSi at 300 K, which is characterized by a metallic contribution below $\omega_{opt}\sim$500 cm$^{-1}$ and by a broad bump peaked at about 700 cm$^{-1}$ and extending up to $\omega_{band}\sim$ 1500 cm$^{-1}$ \cite{commentwband,commentK}. The latter feature is an incipient hybridization induced pseudogap, which evolves in the semiconducting gap excitation at low temperatures (e.g., at 10 K, Fig. 3) \cite{degiorgiepl}. The ratio of the spectral weight belonging to the metallic state with respect to the total one (eq. (1)), including the semiconducting mid-infrared gap, leads to $K_{opt}/K_{band}$ of about 0.4 (Fig. 4), thus placing FeSi in the regime of moderate electronic correlations.

Materials exhibiting colossal magnetoresistive effect are of high current interest in solid state physics, primarily because of potential technological applications. The well-known manganites and Ca-doped EuB$_6$, for which the onset of ferromagnetism is accompanied by a dramatic reduction of the electrical resistivity, have intensively been studied. With magneto-optical investigations on the Eu$_{1-x}$Ca$_x$B$_6$ series \cite{degiorgiprl,broderick} we revealed a phase diagram in support of a scenario based on the double-exchange model \cite{caimiprl}. That model foresees the close proximity of the Fermi level and a magnetization dependent mobility edge so that a ferromagnetic metal-(Anderson) insulator transition (at $T=0$) occurs upon increasing the Ca-content above a critical Ca concentration ($x_{MI}$) of about 0.5. The Ca-doping induces a drift of the mobility edge such that above $x_{MI}$ it goes past the Fermi energy and the spin polarization due to the ferromagnetic transition no longer releases any of the localized states. The Drude weight is thus insensitive to the spin polarization above $x_{MI}$ \cite{caimiprl}. The fingerprint of such a metal-insulator crossover (at $T>0$) was tracked by the spectral weight changes in the Drude component as a function of both magnetic field and temperature \cite{caimiprl}. The optical conductivity $\sigma_1(\omega)$ above the ferromagnetic transition of selected Eu$_{1-x}$Ca$_x$B$_6$, shown in Fig. 3, gives evidence for a metallic component up to $\omega_{opt}\sim$300 cm$^{-1}$, with a broad high frequency tail, extending up to $\omega_{band}\sim$2000 cm$^{-1}$ and due to localized charge carriers. We can calculate $K_{opt}/K_{band}$ (eq. (1)) for selected Ca-dopings by considering the ratio between the effective metallic (Drude) spectral weight (i.e., up to the cut-off energy $\omega_{opt}$) and the overall one encountered in $\sigma_1(\omega)$ below $\omega_{band}$ \cite{commentwband}, thus including the excitations due to the charge carriers located below the mobility edge \cite{commentK}.  Figure 4 shows $K_{opt}/K_{band}$ for $x$=0, 0.3 (i.e., $x\le x_{MI}$) and $x$=0.8 (i.e., $x\ge x_{MI}$). Upon increasing the Ca-content, the optical response across the ferromagnetic metal-insulator transition mimics the evolution which would be expected for a crossover between the moderate and strong electronic correlations' regimes.

\begin{figure}
\center
\includegraphics[width=12cm]{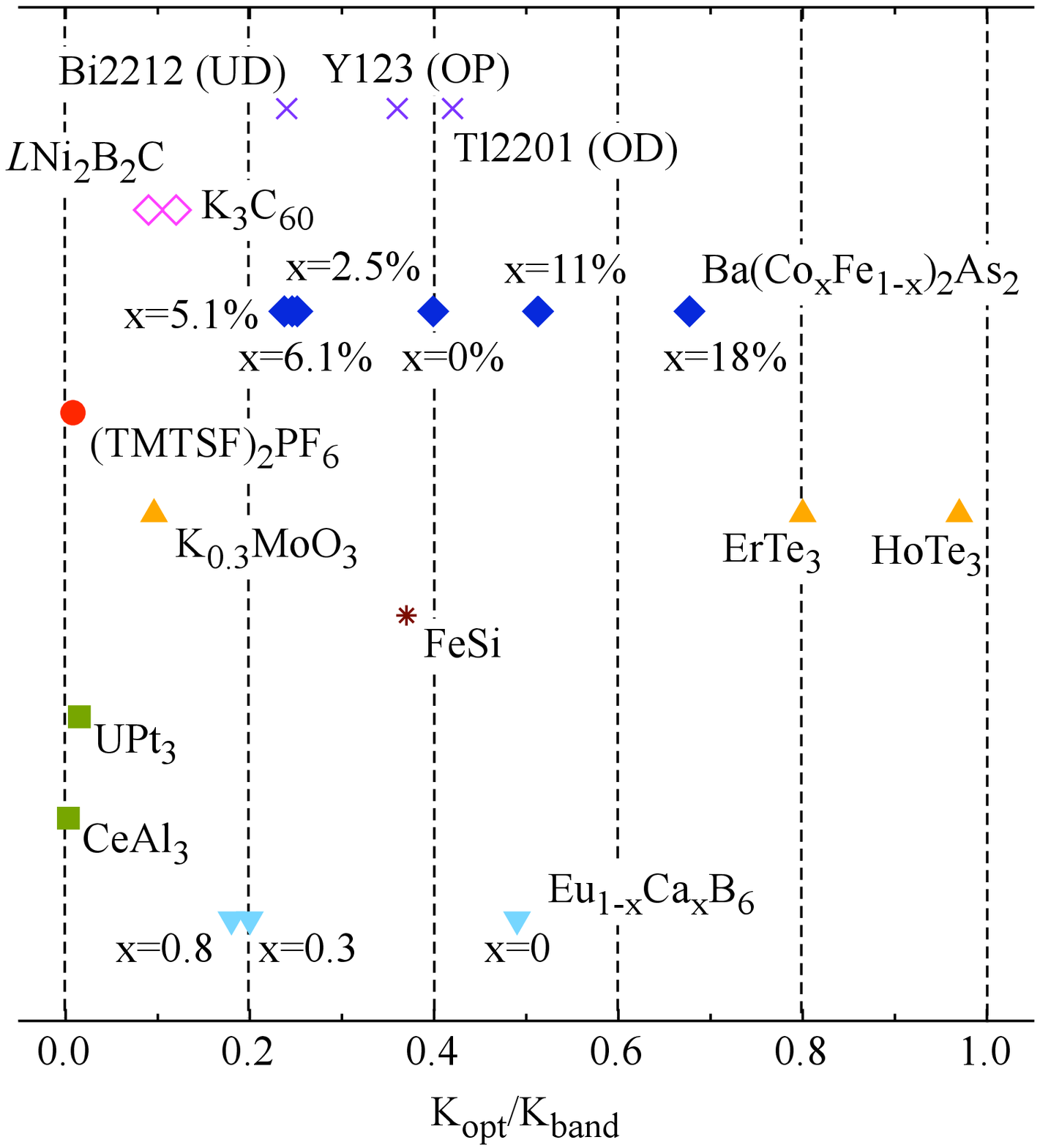}
\caption{(color online) The ratio $K_{opt}/K_{band}$ (eq. (1)) calculated for selected materials, spanning a broad range of the degree of electronic correlations (see text).} \label{correlation}
\end{figure}

We conclude our survey by addressing the broken symmetry ground state due to the formation of a charge-density-wave (CDW) condensate, which derives from the Peierls transition \cite{grunercdw}. The paradigm of CDW forming materials are the quasi-one-dimensional (1D) compounds but electronically driven CDW states were also found in novel two-dimensional (2D) layered compounds \cite{rouxel}. The physics of low-dimensional CDW systems recently experienced a revival of interest. Particularly 2D CDW materials were thoroughly re-investigated, an effort motivated in part by the fact that high-temperature superconductivity in the cuprates may emerge from a peculiar charge ordering in 2D through the tuning of relevant parameters \cite{kivelson}. In Fig. 3 we display $\sigma_1(\omega)$ for a prototype layered CDW system, ErTe$_3$, both in the normal and CDW state \cite{pfuner}. Above the CDW transition temperature ($T_{CDW}$) besides the metallic contribution up to $\omega_{opt}$ there is a broad mid-infrared high energy tail up to $\omega_{band}$ \cite{commentwband}. In quasi-1D CDW materials, like the well-known K$_{0.3}$MoO$_3$, the broad mid-infrared feature clearly appears as a pseudogap-like excitation \cite{schwartz2}. This latter feature, developing above $T_{CDW}$, is ascribed to a precursor CDW gap because of fluctuation effects and evolves into the CDW single-particle excitation below $T_{CDW}$ (Fig. 3). Therefore, $K_{band}$ after eq. (1) results from the total spectral weight, given by the sum of the effective metallic contribution (then defining $K_{opt}$) and the weight encountered in the MIR feature due to the Fermi surface gapping as consequence of the CDW precursor effects. For typical 1D materials, like K$_{0.3}$MoO$_3$, the resulting $K_{opt}$/$K_{band}$ is astonishingly small, of the order of 0.1, while for the 2D rare-earth tritellurides, as ErTe$_3$ or HoTe$_3$, is of about 0.8-1 (Fig. 4). This would imply that the reduction of the electrons' kinetic energy is negligible in 2D materials, as usually expected when electron-phonon coupling is at work. That $K_{opt}$/$K_{band}$ is, on the other hand, substantially reduced in truly chain-like 1D systems possibly originates from an extremely strong electron-phonon coupling, leading to the formation of polarons \cite{perfetti}. The interesting case of 1D CDW materials allows us to broaden the notion of reduced kinetic energy not only as a consequence of electron-electron but also of electron-phonon interaction, thus generalizing the impact of interactions on the charge dynamics as well as ultimately the concept of correlation.

\section{Conclusion}
The present survey over a large wealth of materials provides a rather clear-cut evidence for a direct relationship between the strength of interaction (electron-electron as well as electron-phonon) and the experimentally evinced reduction of the (Drude) metallic spectral weight in the excitation spectrum. The total spectral weight encountered in the optical conductivity up to the onset of the electronic interband transitions turns out to be usually redistributed into a well-distinct Drude resonance and a so-called incoherent part, identified throughout the paper with the generic MIR band concept. We consequently figure out that the stronger is the Coulomb repulsion or the electron-phonon coupling, the stronger is the deviation of the charge-dynamics from a simple band-metal response, pushing the materials into the strongly correlated regime which is comparable with a Mott-insulating or a polaron's dominated state, respectively. 

Since the origin of interactions may be very diverse, the physical meaning of the MIR feature, due to the incoherent part of $\sigma_1(\omega)$, is also very different from system to system. Similarly, the reduced Drude weight, as effective integrated quantity, may originate from changes of the charge density as well as of the mass enhancement of the itinerant carriers, depending from the physics involved in the investigated material. We made the assumption that the effective (reduced) Drude weight should be compared to the total one encountered in $\sigma_1(\omega)$ below the interband transitions. However, it remains to be seen, how the total weight up to $\omega_{band}$ deviates from the LDA expectation in the nearly-free electron limit. Nevertheless, we believe that our definition of the degree of electronic correlations from spectral weight argument in terms of the $K_{opt}/K_{band}$ ratio would be little influenced by those deviations, thus representing a truly experimental and reliable method towards its estimation.

\begin{acknowledgments}
The author wishes to thank A. Lucarelli, M. Dressel, L. Benfatto, D. Basov, M. Qazilbash and A. Chubukov for fruitful discussions. This work has been supported by the Swiss National Foundation for the Scientific Research
within the NCCR MaNEP pool. 
\end{acknowledgments}


\begin{thebibliography}{99}
\bibitem{commentcorr} That superconductivity in the iron-pnictides emerges in close proximity of an antiferromagnetic (AF) ground state implies the important role of AF fluctuations for the pairing mechanism of all itinerant charge carriers. Both Drude terms are affected by the superconducting transition \cite{wu} and it is consequently legitimate to consider the total spectral weight of both of them (i.e., involving all ungapped charge carriers) for the definition of the optical kinetic energy. For low Co-dopings (particularly for $x\le0.025$), the SDW instability and the resulting Fermi surface gapping at low temperatures within the SDW state obviously leads to a reshuffling of spectral weight, thus affecting both quantities $K_{opt}$ and $K_{band}$ in eq. (1). Therefore, $K_{opt}/K_{band}$ after eq. (1) for the underdoped compounds ($x\le$0.025) is smaller at $T<T_{SDW}$ than above. Nevertheless, it turns out that $K_{band}/K_{opt}$ both at high and low temperature behaves similarly as a function of Co-doping, so that both estimations converge with increasing doping-content and are basically identical for $x\ge$0.11. Thus, the proposed average value in Fig. 2 suitably gives an overall trend for the degree of electronic correlations.
\bibitem{commentwband} By Drude response, we intend the truly metallic component in $\sigma_1(\omega)$, generally reproduced by a simple Drude term \cite{wooten,grunerbook}. Such a Drude term is usually well distinct up to a cut-off frequency $\omega_{opt}$. On the other hand, $\omega_{band}$ denotes the onset of the electronic interband transitions and it is generally defined as the zero-crossing energy of the imaginary part of the complex optical conductivity \cite{basov}.
\bibitem{commentK} $K_{opt}\sim\int_{0}^{\omega_{opt}}\sigma_1(\omega)d\omega$ and $K_{band}\sim\int_{0}^{\omega_{band}}\sigma_1(\omega)d\omega$ (Fig. 3). Alternatively, one can exploit the phenomenological Drude-Lorentz fit \cite{wooten,grunerbook}, so that $K_{opt}\sim\omega_p^2$ and $K_{band}\sim\omega_p^2+\sum_{j}S_j^2$ ($\omega_p$ being the plasma frequency of the effective Drude term shaping $\sigma_1(\omega)$ up to $\omega_{opt}$ and $S_j^2$ the mode strength of the $j-$Lorentz h.o.'s contributing to $\sigma_1(\omega)$ up to $\omega_{band}$). Details to the fits and related parameters can be evinced for each material from the quoted references.
\bibitem{commentmetal} The $\sigma_1(\omega)$ spectrum of conventional metals, like Au and Al, can be reproduced by a simple Drude response, the integral of which leads to values of the plasma frequency $\omega_p\sim\sqrt{n/m_e}$, totally in agreement with the theoretical expectation given by the charge carrier concentration ($n$) from chemical counting and the free electron mass ($m_e$) \cite{grunerbook}.
\end{thebibliography}
\end{document}